# Enhancement of heat and mass transfer by herringbone microstructures in a simple shear flow

Yanxing Wang[1,*], Hui Wan[2], Tie Wei[3], John Abraham[4]

[1] Department of Mechanical and Aerospace Engineering, New Mexico State University, Las Cruces, NM 88011, USA

[2] Department of Mechanical and Aerospace Engineering, University of Colorado Colorado Springs, Colorado Springs, CO 80918, USA

[3] Department of Mechanical Engineering, New Mexico Institute of Mining and Technology, Socorro, NM 87801, USA

[4] Department of Mechanical Engineering, University of St. Thomas, St. Paul, MN 55105, USA

**Abstract**

The heat and mass transfer characteristics in a simple shear flow over staggered herringbone structures are numerically investigated with the lattice Boltzmann method. Two flow motions are identified. The first is a spiral flow oscillation above the herringbone structures that advects heat and mass from the top plane to herringbone structures. The second is a flow recirculation in the grooves between herringbone ridges that advects heat and mass from the area around herringbone tips to the side walls of herringbone ridges and the bottom surfaces. These two basic flow motions couple together to form complex transport mechanisms. The results show that when advective heat and mass transfer takes effect at relatively larger Reynolds and Schmidt numbers, the dependence of the total transfer rate on the Schmidt number follows a power law, with the power being the same as that in the Dittus-Boelter equation for turbulent heat transfer. As Reynolds number increases, the dependence of the total transfer rate on Reynolds number also approaches a power law, and the power is close to that in the Dittus-Boelter equation.

## 1. Introduction

Microfluidics have received enormous attention in the past decades due to the advancements of micro fabrication technologies, and have been broadly used in the fundamental and applied studies of physical, chemical, and biological processes [1-3]. Due to the small scale of micro channels, the flow is highly laminar, and diffusion is usually the primary mechanism for species transport and mixing. This diffusion takes place on much longer length and time scales than does convection. Chaotic advection can be employed with embedded micro-structures on the surface of micro-channels to generate a transverse flow. The transverse flow advects dissolved substances over the cross section and significantly enhances the mixing and transport efficiency [4, 5]. One of the most efficient chaotic micromixer is the Staggered Herringbone Mixer (SHM) developed by Stroock et al. [4] The repeated patterns of grooves on the inner surface of the SHM creates helical motions of the fluid in the microchannels, thus providing a mixing and transport mechanism through transversal advection.

Due to the simple fabrication and high mixing efficiency of the SHM, numerous studies have been conducted, including the analysis of geometric effects on mixing efficiency [6-8], the design and optimization of micro-mixers with different applications [9-13], and development of new designs for specific applications [14, 15]. However, most of these studies focus on the mixing of two species in T-type SHM mixers in which two flows with different species enter the mixer through two inlet branches of the T shape [16]. The helical pattern of the flow generated by staggered herringbone structures suggests that the flow provides a transverse advection which might enhance the heat and mass transport between the bulk flow and the bottom wall. Several studies have shown that the SHM can significantly enhance the convective and boiling heat transfer inside microchannels [17, 18]. This promising mechanism might have broad applications in compact heat exchangers and microelectromechanical systems (MEMS), and could also be extended to the areas of chemistry, biology, and those involving micro-scale mass transfer. Surprisingly, herringbone-inspired microstructures have gained little attention for its heat and mass transfer capacity up to date.

---

[*] Corresponding author.
Email: yxwang@nmsu.edu

For a flow in a micro-channel embedded with staggered herringbone structures, a boundary layer with reduced velocity components and a helical flow pattern is developed over the structures. The generation of the helical flow pattern is directly related to the shear rate and thickness of the boundary layer. In this study, we consider a fully developed simple shear flow confined by two infinitely large planes with staggered herringbone structures embedded on the lower surface. With lattice Boltzmann method, the heat and mass transfer from the top plane to the bottom plane are numerically investigated. This study aims to identify the mechanisms for the enhancement of heat and mass transfer by the transverse flow advection over the staggered herringbone structures. This study also investigates the dependence of transfer efficiency on the influencing parameters, such as, shear rate, Prandtl number, and Schmidt number. The rest of the paper is organized as follows. A detailed description of the physical model is presented in Section 2. The numerical method is described in Section 3. The results are analyzed in Section 4, and the concluding remarks are given in Section 5.

## 2. Physical Model

As shown in Fig. 1, we model a laminar incompressible simple shear flow enclosed by two infinitely large parallel planes, with staggered herringbone structures embedded on the bottom plane. To generate a simple shear flow, the top plane moves at a constant velocity $U_0$, and the bottom plane is fixed. The distance between two planes, $H$, is three times the height of the herringbone ridges, $h$. The width of the herringbone ridges, $w$, is half of $h$, and the spacing between neighboring ridges $\delta$ is the same as $h$. The angle between the ridges of herringbone structures and the streamwise axis, $\theta$, is fixed at 45°. The regular arrangement of the herringbone ridges forms the periodicity of the surface geometry in both streamwise and spanwise directions. The streamwise dimension of each unit containing a complete herringbone element can be calculated as $L_x = (\delta + w)/sin\theta$. The spanwise dimension of each unit $L_y$ is $2L_x$. With these geometric specifications, the flow evolution is exclusively determined by the shear Reynolds number defined based on the effective shear rate $S$ and the flow passage height $H - h$. The effective shear rate is defined as:

$$S \equiv \frac{U_0}{H-h} \tag{1}$$

and the shear Reynolds number is defined as

$$Re_S \equiv \frac{S(H-h)^2}{v} \tag{2}$$

where $v$ is the kinematic viscosity of the working fluid. In this study, the shear Reynolds number ranges from 20 to 200. Below that range, the diffusive transfer plays a dominant role.

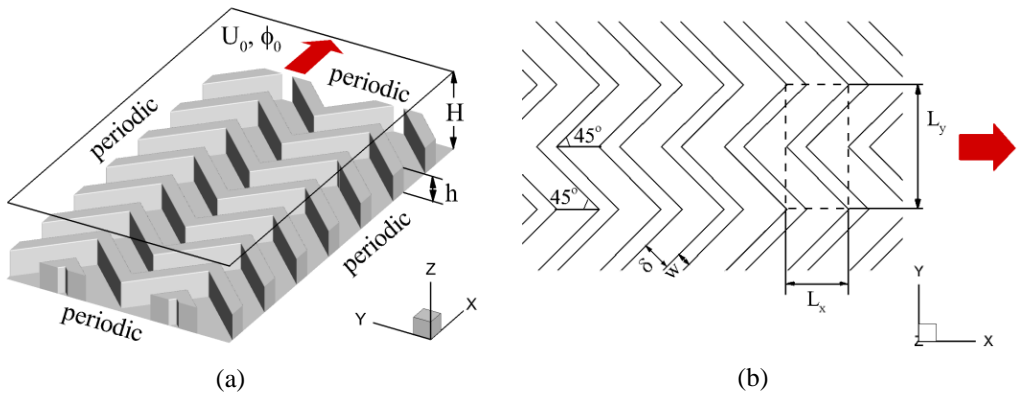

Fig. 1. Physical model of heat and mass transfer in a simple shear flow with herringbone structures on bottom wall. (a) 3D view, and (b) top view. The dashed rectangle indicates the domain of numerical simulation.

We will develop a general understanding of the enhancement of heat and mass transfer induced by herringbone structures, so the shear Reynolds numbers considered here are higher than those for microfluidic systems. In this study, the temperature and the concentration of dissolved species are modeled as a passive scalar, released at the top plane and absorbed at both the bottom plane and the surfaces of herringbone ridges. For purpose of generalization, a nondimensional scalar concentration is utilized, which is fixed at 1 at the top plane, and 0 at the bottom plane and the surfaces of herringbone ridges. The Schmidt number for scalar diffusion in this study ranges from 1 to 50. The characteristics of scalar transfer from top plane to bottom surfaces are investigated in both diffusion dominant and advection dominant ranges.

The height of the herringbone ridges $h$ and the velocity of top plane $U_0$ are used as the characteristic length and velocity to normalize the spatial coordinates and flow velocity,

$$\tilde{x} = x/h, \tag{3}$$

$$\tilde{\mathbf{u}} = \mathbf{u}/U_0, \tag{4}$$

In normalized form, the height of herringbone ridges is $\tilde{h} = 1$ and the velocity of upper plane is $\tilde{U}_0 = 1$.

## 3. Numerical Methods

In this study we developed a 3-D numerical model based on the lattice-Boltzmann method (LBM) to model continuum level meso and micro scale incompressible fluid flow with complex surface geometries. The dependent variable is the particle distribution function $f_\alpha(\mathbf{x}, t)$ that quantifies the probability of finding an ensemble of molecules at position $\mathbf{x}$ with velocity $\mathbf{e}_\alpha$ at time $t$ [19-21]. In three dimensions, the velocity vector $\mathbf{e}$ can be discretized into 15, 19 or 27 components (referred to as D3Q15, D3Q19 and D3Q27) [19]. Here we applied the D3Q15 approach largely to minimize computational load, with the recognition that the Reynolds number is relatively low. LBM is well suited to the present problem because of its powerful capability in dealing with complex geometries and its highly parallelizability.

The Boltzmann equation discretized on a lattice with the BGK form of collision operator is given for single-phase flow by Chen & Doolen [20], and Wang et al. [21],

$$f_\alpha(\mathbf{x} + \mathbf{e}_\alpha \delta_t, t + \delta_t) - f_\alpha(\mathbf{x}, t) = -\frac{1}{\tau}\left(f_\alpha(\mathbf{x}, t) - f_\alpha^{eq}(\mathbf{x}, t)\right) \tag{5}$$

where $\mathbf{e}_\alpha$ is the elementary velocity vector in direction $\alpha$, $\tau$ is the relaxation time and $f_\alpha^{eq}$ is the equilibrium distribution function in direction $\alpha$.

$$f_\alpha^{eq}(\mathbf{x}, t) = w_\alpha \rho \left(1 + \frac{\mathbf{e}_\alpha \cdot \mathbf{u}}{c_s^2} + \frac{(\mathbf{e}_\alpha \cdot \mathbf{u})^2}{2c_s^4} - \frac{\mathbf{u} \cdot \mathbf{u}}{2c_s^2}\right) \tag{6}$$

The symbol $w_\alpha$ is the weighting coefficient, $c_s$ is the sound speed in the lattice, and $\mathbf{u}$ is the fluid velocity. The right-hand side of Eq. (5) describes the mixing, or collision of molecules that locally drives the flow to an equilibrium particle distribution, $f_\alpha^{eq}(\mathbf{x}, t)$. Macroscopic variables such as density $\rho$ and velocity $\mathbf{u}$ are calculated from the moments of the distribution functions,

$$\rho(\mathbf{x}, t) = \sum_\alpha f_\alpha(\mathbf{x}, t), \qquad \rho(\mathbf{x}, t)\mathbf{u}(\mathbf{x}, t) = \sum_\alpha f_\alpha(\mathbf{x}, t)\mathbf{e}_\alpha \tag{7}$$

As is common, we apply the BGK model for this collision process through which the distribution functions $f_\alpha(\mathbf{x}, t)$ relax towards $f_\alpha^{eq}(\mathbf{x}, t)$ with a single lattice relaxation time scale,

$$\tau = (6v/(c\delta x) + 1)/2 \quad \text{and} \quad c = \delta x/\delta t. \tag{8}$$

In the traditional treatment of solid boundaries, a solid wall is assumed to be located half way between the lattice nodes, the molecules traveling toward the wall are bounced back at the wall and return to the same node. Consequently, the distribution function in the left direction is the same as that in the right direction before streaming. The accuracy of this scheme is only 1st order. In the present study, we used the scheme with 2nd order of accuracy proposed by Lallemand and Luo [22]. This scheme is based on a simple bounce-back treatment and interpolations. If the distance from the first fluid node from the solid boundary, $q$, is less than half the lattice space, the computational quantities are interpolated before propagation and bounce-back collision. If $q$ is greater than the half lattice space, interpolation is conducted after propagation and bounce-back collision. The momentum exerted by the moving boundary is given by these terms, which were first advanced by Ladd [23].

The "moment propagation method" developed by Frenkel and Ernst [24], Lowe and Frenkel [25], and Merks et al. [26] is used to solve for the temperature. In this method, a scalar quantity $T$ is released in the lattice and a scalar concentration field variable is propagated at the continuum level for each scalar using the particle distribution function $f_\alpha(\mathbf{x}, t)$.

$$T(\mathbf{x}, t + \delta t) = \sum_\alpha P_\alpha(\mathbf{x} - \mathbf{e}_\alpha \delta t, t + \delta t) + \Lambda T(\mathbf{x}, t) \quad (9)$$

where

$$P_\alpha(\mathbf{x} - \mathbf{e}_\alpha \delta t, t + \delta t) = \left[\frac{\hat{f}_\alpha(\mathbf{x} - \mathbf{e}_\alpha \delta t, t_+)}{\rho(\mathbf{x} - \mathbf{e}_\alpha \delta t, t)} - w_\alpha \Lambda\right] T(\mathbf{x} - \mathbf{e}_\alpha \delta t, t) \quad (10)$$

$\hat{f}_\alpha$ is the post-collision distribution function,

$$\Lambda = 1 - 6\,\varepsilon/c\delta x \quad (11)$$

and $\varepsilon$ is the scalar diffusivity.

We first selected the cases with $Re_S = 200$ and ran the simulation over 5 structural elements in streamwise direction. The results demonstrated that the spatial periodicity of flow pattern is the same as the streamwise length of the structural element. Therefore, in the large-scale systematic study we only simulated the flow within the cuboid domain including one structural element as indicated by the dashed lines in Fig. 1(b). Periodic conditions in the streamwise and spanwise directions were used. The total number of computational grid cells is $100 \times 200 \times 140$, and 24 grid cells are used to resolve the ridge thickness of the herringbone structure. The computational domain is decomposed into 56 subdomains. The Message Passing Interface (MPI) technique was utilized to conduct parallel computing. The analysis was carried out when the flow entered into the stationary state.

The LBM solver has been extensively validated in previous studies and more details are described in Wang et al. [21, 27-29]. To examine the grid sensitivity of the results, simulations were conducted with two different sets of grids corresponding to different resolutions. The number of grid points over one ridge thickness of the herringbone structures is 24 and 36, respectively. Figure 2 shows the profiles of the horizontally averaged streamwise velocity and scalar concentration for $Re_S = 200$ and $Sc = 50$, respectively. Excellent agreement between the coarse and fine grid is obtained. The maximum deviation between the two grids is less than 2%. With the excellent agreement between the two meshes, mesh independence is assured and hereafter, results will be obtained using the coarse one.

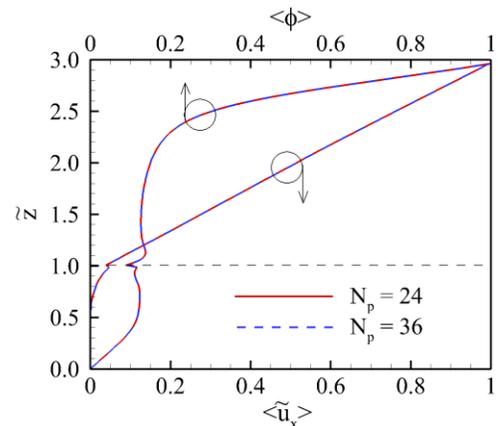

Fig. 2. Effect of grid resolution on horizontally averaged streamwise velocity $\langle \tilde{u}_x \rangle$ and scalar concentration $\langle \phi \rangle$ at $Re_S = 200$ and $Sc = 50$. $N_p$ is the number of grid points over one ridge thickness of herringbone structures.

## 4. Results and Discussion

Scalar transport from the top plane to the bottom plane relies on both diffusion and advection induced by the herringbone structures. Advective transport is dominant when the Reynolds and Schmidt numbers are large. Our focus is on the identification of the mechanism for the enhancement of scalar transfer caused by herringbone structures. $Re_S = 200$ was selected as a prototypical example to analyze the flow behavior and scalar transfer characteristics.

Figure 3 shows the patterns of flow characteristics for $Re_S = 200$. Taking advantage of the periodicities in the streamwise and spanwise directions, only the flow within the cuboid domain including a single structural element needed to be simulated. The patterns containing multiple elements shown in the figure were generated by concatenating the patterns of one element. The typical streamlines over herringbone structures are shown in Fig. 3(a). In the steady state, the streamlines coincide with the trajectories of fluid particles, which illustrate the path of scalar transport by flow advection. The color represents the level of scalar concentration at $Sc = 50$. The red and blue colors indicate higher and lower concentrations, respectively. The streamline pattern shows how the flow transports a scalar from the bulk flow to the surfaces in the herringbone grooves for absorption: the fluid with higher scalar concentration flows downward in the streamwise strip area along the backward connecting ends (BCE) of the herringbone structures. The flow is entrapped in the grooves at the BCE of the grooves where it recirculates within the grooves before flowing upward and leaving the grooves with lower scalar concentration at the forward connecting ends (FCE) of the grooves. Subsequently, the fluid flows upward in the streamwise strip along the FCE of the structures. This flow motion provides the fluid a greater contact with the surfaces of the herringbone ridges and the bottom surfaces between herringbone ridges and enhances the heat and mass exchange between the bulk flow and the surfaces with complex geometry.

It is well known that the vertical velocity component provides a mechanism for heat and mass transport in the vertical direction by advection. Figure 3(b) shows the iso-surfaces of the vertical velocity at $\tilde{u}_z = \pm 0.005$. Since the flow is disturbed in the lower region, the magnitude of $\tilde{u}_z$ is larger than that in the upper region. As shown in the figure, above every BCE of the grooves where fluid is entrapped in the grooves, a region with negative $\tilde{u}_z$ is generated. The iso-surfaces of $\tilde{u}_z$ are consistent with the streamline patterns in Fig. 3(a). That is, in the streamwise strip area above the BCE of the grooves, the flow goes downward, and in the streamwise strip area above the FCE of grooves, the flow goes upward. Figure 3(c) shows the corresponding 2D iso-contours of vertical velocity averaged over the streamwise coordinate, $\langle \tilde{u}_z \rangle_x$, on the cross section ($(\tilde{y}, \tilde{z})$ plane). The variable averaged over the streamwise coordinate is defined as

$$\langle \tilde{\alpha} \rangle_x (\tilde{y}, \tilde{z}) \equiv \frac{1}{\tilde{L}_x} \int_{\tilde{L}_x} \tilde{\alpha}(\tilde{x}, \tilde{y}, \tilde{z}) d\tilde{x} \qquad (12)$$

where $\tilde{\alpha}(\tilde{x}, \tilde{y}, \tilde{z})$ is the variable of interest, and $\tilde{L}_x$ is the length of a complete herringbone element in the streamwise direction. The iso-surfaces of $\tilde{u}_z$ in Fig. 3(b) and iso-contours of $\langle \tilde{u}_z \rangle_x$ in Fig. 3(c) describe the patterns of vertical velocity of the fluid that has been disturbed by the herringbone structures. Compared with the velocity at the top plane, the vertical velocity around the herringbone structures is much smaller. This implies that the pitch of the helix is much larger than the height of flow passage $(H - h)$. In the upper region, it is well known that each unit of staggered herringbone structures generates a pair of counter-rotating tubular eddies extending in the streamwise direction and the fluid particle flows along a helical trajectory in the tubular eddies[2, 4]. The helical pattern can be inferred from the 2D streamlines, based on the streamwise averaged velocity components, $\langle \tilde{u}_y \rangle_x$ and $\langle \tilde{u}_z \rangle_x$, as shown in Fig. 3(d). The closed streamlines define the recirculating motions in the $(\tilde{y}, \tilde{z})$ plane of the helix. This continuous upward and downward flow motion enhances the scalar transport in the vertical direction.

To quantitatively compare the flow characteristics under different conditions, we define the horizontally averaged quantities as,

$$\langle \tilde{\beta} \rangle(\tilde{z}) \equiv \frac{1}{A} \int_A \tilde{\beta}(\tilde{x}, \tilde{y}, \tilde{z}) d\tilde{x} d\tilde{y} \qquad (13)$$

where $\tilde{\beta}(\tilde{x}, \tilde{y}, \tilde{z})$ is the quantity of interest that varies with spatial coordinates, and $A$ is the horizontal area occupied by fluid at different vertical locations in a herringbone element. The averaged variable $\langle \tilde{\beta} \rangle$ is a function of the vertical coordinate $\tilde{z}$ only. The horizontally averaged vertical velocity $\langle \tilde{u}_z \rangle$ represents the net flux of fluid (flow per unit area) through the horizontal plane. As a result of mass conservation, $\langle \tilde{u}_z \rangle$ is zero for the present problem. To quantify the wall-normal advective transfer capability, we define the effective upward and downward velocity as

$$\tilde{u}_z^+ \equiv \begin{cases} \tilde{u}_z, & if\ \tilde{u}_z > 0 \\ 0, & if\ \tilde{u}_z < 0 \end{cases} \quad and \quad \tilde{u}_z^- \equiv \begin{cases} 0, & if\ \tilde{u}_z > 0 \\ \tilde{u}_z, & if\ \tilde{u}_z < 0 \end{cases} \tag{14}$$

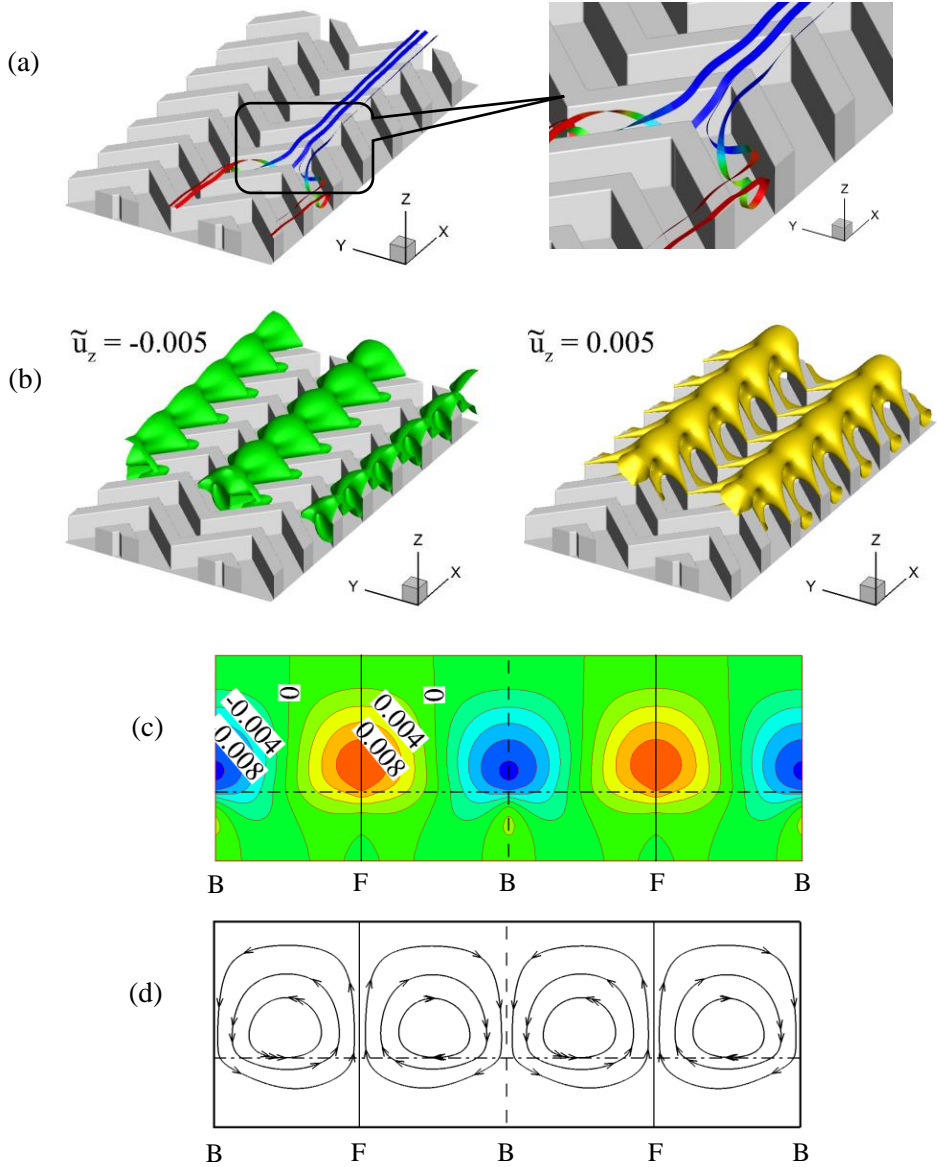

Fig. 3. Flow patterns induced by herringbone structures in a simple shear flow at $Re_S = 200$. (a) typical streamtraces entraped into the grooves of herringbone structures, (b) iso-surfaces of vertical velocity at $\tilde{u}_z = \pm 0.005$, (c) 2D contours of vertical velocity averaged over streamwise coordinate $\langle \tilde{u}_z \rangle_x$, and (d) streamline patterns based on the streamwise averaged velocity components $\langle \tilde{u}_y \rangle_x$ and $\langle \tilde{u}_z \rangle_x$ on a cross section. The letters 'B' and 'F' in (c) and (d) indicate the positions of backward and forward connecting points of the grooves.

In this study, we use $\tilde{u}_z^-$ to deal with the heat and mass transfer from the bulk flow to the bottom surface.

Figure 4 plots the profiles of horizontally averaged streamwise velocity $\langle\tilde{u}_x\rangle$ and the effective downward velocity $\langle\tilde{u}_z^-\rangle$ at $Re_S = 20$ and 200. The horizontal dashed line indicates the position of the ridge tips of the herringbone structures. The curves of $\langle\tilde{u}_x\rangle$ shows little difference between $Re_S = 20$ and 200. Yet the curves of $\langle\tilde{u}_z^-\rangle$ do exhibit apparent differences. This suggests that the influence of Reynolds number primarily occurs with the vertical velocity component. Above the ridge tips ($\tilde{u} > 1$), $\langle\tilde{u}_x\rangle$ decreases roughly linearly from 1 at the upper plane to a small value close to 0 at the ridge tips. In the grooves ($\tilde{u} < 1$), $\langle\tilde{u}_x\rangle$ remains a small value.

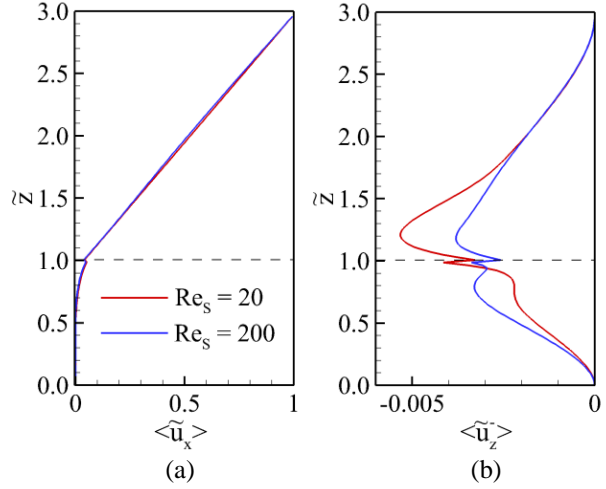

Fig. 4. Profiles of horizontally averaged velocity components. (a) streamwise velocity $\langle\tilde{u}_x\rangle$, and (b) effective downward velocity $\langle\tilde{u}_z^-\rangle$. The horizontal dashed lines indicate the tips of herribone ridges.

The curves of $\langle\tilde{u}_z^-\rangle$ have two peaks, one is above the herringbone ridge tips and the other is below the tips, corresponding to the consecutive upward and downward flow motions in the helix in the upper region and the flow recirculation in the grooves, respectively. In the vicinity of the herringbone structures, the velocity magnitude is larger, so the upper peak is close to the herringbone tips in the curve. The magnitude of $\langle\tilde{u}_z^-\rangle$ largely indicates the advective scalar transport capability of the two flows. As shown in the figure, at the upper peak the value of $|\langle\tilde{u}_z^-\rangle|$ for $Re_S = 20$ is larger than that for $Re_S = 200$. Yet at the lower peak, the value of $|\langle\tilde{u}_z^-\rangle|$ for $Re_S = 20$ is smaller than that for $Re_S = 200$. Hence, in the upper region, the flow at $Re_S = 20$ has a greater scalar transport capability, from the upper bulk flow to the herringbone top surfaces and to the recirculating flow in the grooves.

In the grooves, the situation is just the opposite. The flow at $Re_S = 200$ shows a greater scalar transport capability to the bottom surface. This is because in the upper region the greater viscous effect at smaller Reynolds number enables the disturbance of herringbone structures propagate a longer distance in the flow, and in the grooves the flow recirculation are more active at larger Reynolds numbers. The overall heat and mass transfer process depends on the coupling of these two mechanisms through a complex relationship. Generally, the flows at larger Reynolds numbers have a stronger advective scalar transport capability.

The scalar transport also depends on the Schmidt number. For heat transfer, the corresponding characteristic number is the Prandtl number. Here we use the flow at $Re_S = 200$ to analyze the effect of the Schmidt number. Figure 5 shows the patterns scalar concentration for $Sc = 1$ and 50. At $Sc = 1$, the diffusion effect is so strong that the diffusive scalar transport dominates over the advective scalar transport. The 3D iso-surfaces and 2D contours on two typical cross sections shown in Fig. 5(a) suggest that the scalar concentration ϕ changes roughly linearly with the vertical coordinate in the flow above the herringbone structure. The herringbone ridges only cause a slight disturbance to the distribution of ϕ. In the grooves, ϕ remains at a low level. This pattern means that the majority of the scalar is absorbed at the top surface. This further means that the advection of the recirculating flow is the primary means for scalar transport in the grooves, and the reduced advective scalar transport of the recirculating flow at lower $Sc$ cannot effectively transport a scalar to the bottom surfaces in the grooves.

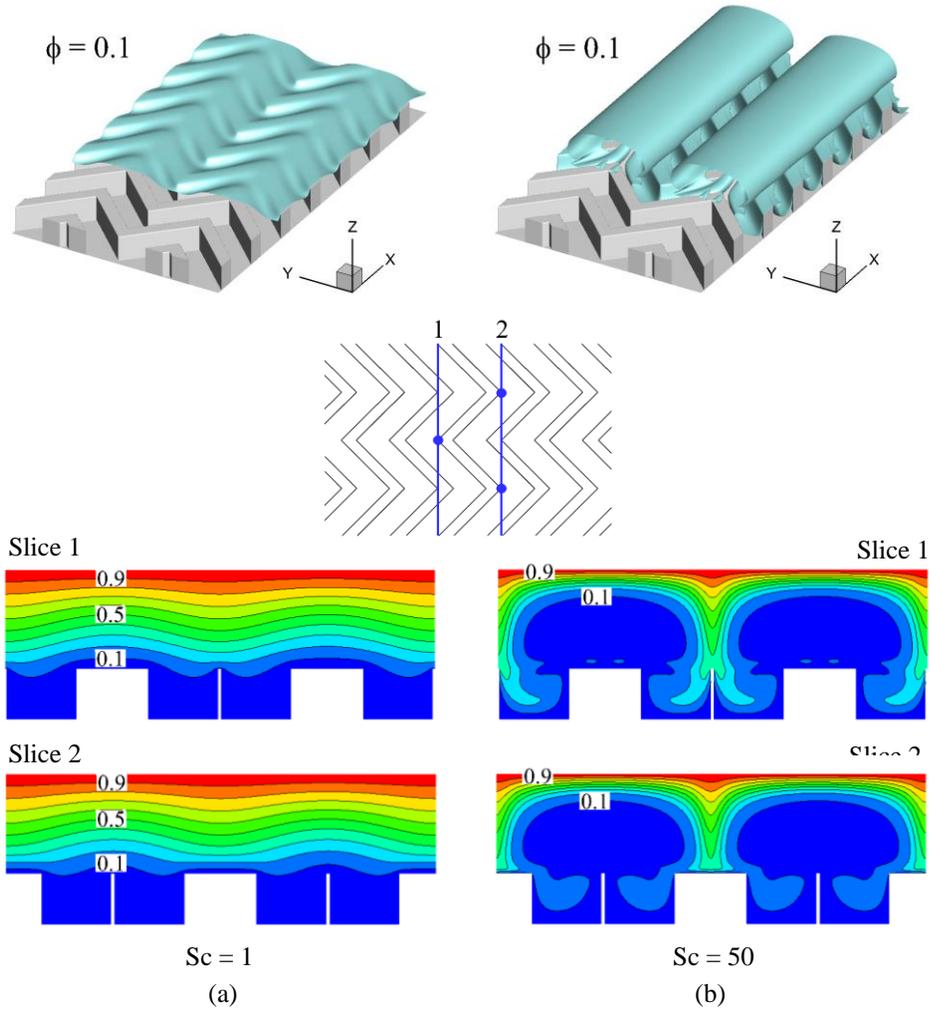

Fig. 5. Patterns of scalar concentration ϕ at $Re_S = 200$. (a) $Sc = 1$, and (b) $Sc = 50$. The slices are selected at the cross sections through the backward and forward connecting points of the grooves as shown in the sketch.

At $Sc = 50$, the diffusive scalar transport is reduced and the advective transport becomes dominant. The comparison of Fig. 5(b) with Fig. 3 suggests that the flow goes upward with a lower scalar concentration in the strip area along the FCE of the grooves and downward with a higher scalar concentration in the strip area along the BCE of the grooves. In the grooves, the recirculating flow helps to transport the scalar to the bottom surfaces and the side walls of the herringbone ridges. In the top region, the flow acquires the scalar from the top plane through scalar diffusion. These consecutive steps articulate the whole process of scalar transport from the top plane to the bottom surfaces and ridge surfaces. The comparison between Figs. 5(a) and 5(b) confirms the conclusion that the advective scalar transfer primarily takes effect at larger Schmidt numbers.

Figure 6 shows the advective scalar transport flux in vertical direction $\tilde{u}_z \phi$ for $Sc = 1$ and 50 at $Re_S = 200$. The 3D iso-surfaces and 2D iso-contours of $\tilde{u}_z \phi$ are consistent with the patterns of vertical velocity $\tilde{u}_z$ shown in Fig. 3. Along the FCE of the grooves, the values of $\tilde{u}_z \phi$ are positive, indicating an upward advective scalar flux. Along the BCE of the grooves, the values of $\tilde{u}_z \phi$ are negative, indicating a downward advective scalar flux. At $Sc = 1$, the stronger diffusion effect reveals tubular structures in the iso surfaces that extend in the streamwise direction on each row of the herringbone ridges. At $Sc = 50$, the enhanced advection effect creates complex fine structures on the iso-surfaces. The comparison of the 2D contours suggest that at $Sc = 1$ the value of $\tilde{u}_z\phi$ in the upward flow is larger, yet the absolute value of $\tilde{u}_z\phi$ in the downward flow is smaller compared with the values at $Sc = 50$. This implies the net efficiency of advective scalar transport from the top plane to the bottom region is lower at $Sc = 1$. As a result, the total scalar flux is lower.

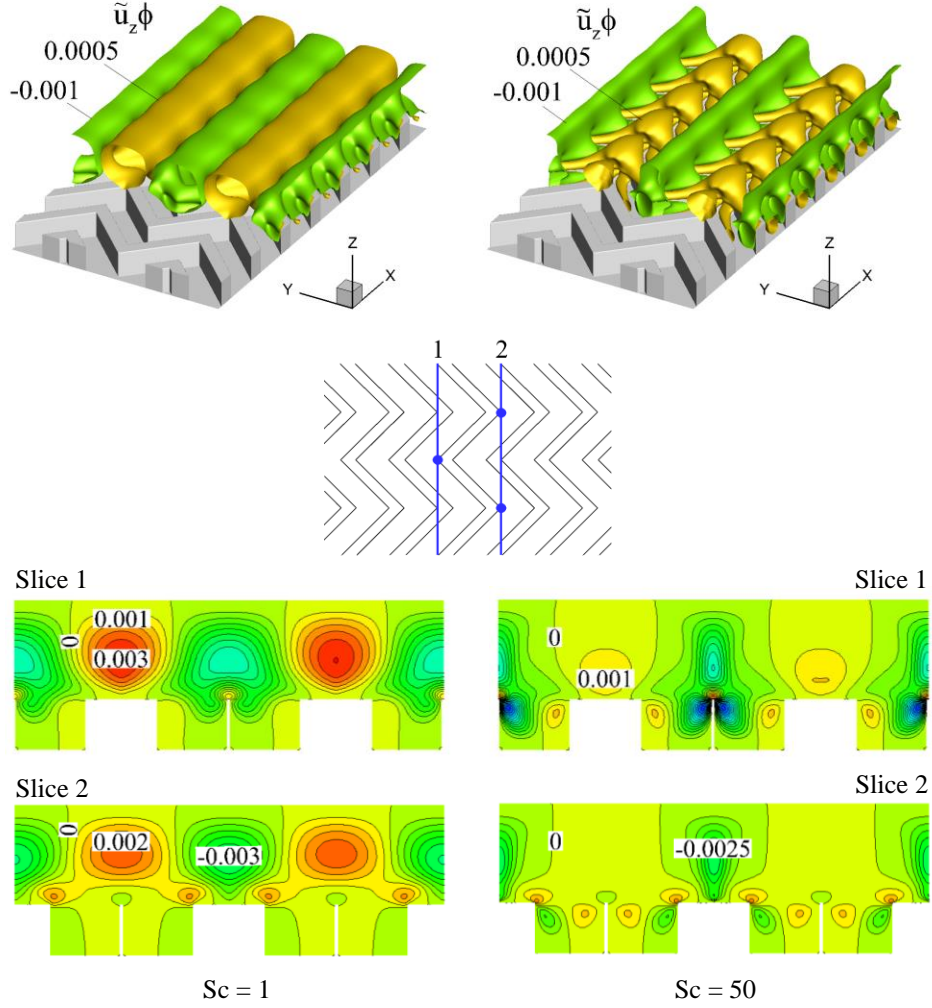

Fig. 6. Patterns of advective scalar flux in vertical direction $\tilde{u}_z\phi$ at $Re_S = 200$. (a) $Sc = 1$, and (b) $Sc = 50$. The slices are selected at the cross sections through the backward and forward connecting points of the grooves as shown in the sketch.

The scalar transport from the top plane to the bottom region relies on both scalar diffusion and flow advection induced by the herringbone structures. In the nondimensional form, the diffusive and advective scalar fluxes, $\tilde{q}_{dif}(\tilde{z})$ and $\tilde{q}_{adv}(\tilde{z})$, are calculated as,

$$\tilde{q}_{dif}(\tilde{z}) = \frac{1}{Pe}\langle\frac{\partial\phi}{\partial\tilde{z}}\rangle \quad (15)$$

$$\tilde{q}_{adv}(\tilde{z}) = -\langle\tilde{u}_z\phi\rangle \quad (16)$$

where $Pe$ is the Peclet number. The total flux $\tilde{q}_{tot}$ is the sum of $\tilde{q}_{dif}(\tilde{z})$ and $\tilde{q}_{adv}(\tilde{z})$,

$$\tilde{q}_{tot} = \frac{1}{Pe}\langle\frac{\partial\phi}{\partial\tilde{z}}\rangle(\tilde{z}) - \langle\tilde{u}_z\phi\rangle(\tilde{z}) \quad (17)$$

Both $\tilde{q}_{dif}(\tilde{z})$ and $\tilde{q}_{adv}(\tilde{z})$ are functions of the vertical coordinate $\tilde{z}$, yet the total scalar flux $\tilde{q}_{tot}$ is independent of $\tilde{z}$ in the steady state due to the conservation of scalar flux.

To quantitatively compare the scalar transport characteristics, Figs. 7 and 8 show the profiles of the horizontally averaged scalar concentration $\langle\phi\rangle$, the gradient of scalar concentration in the vertical direction $\langle\partial\phi/\partial\tilde{z}\rangle$, and the ratio of advective scalar flux to total scalar flux $|\langle\tilde{u}_z\phi\rangle|/\tilde{q}_{tot}$ for various $Sc$ at $Re_S = 20$ and 200, respectively. $\langle\partial\phi/\partial\tilde{z}\rangle$ quantifies the diffusive scalar transport in the vertical direction, and $|\langle\tilde{u}_z\phi\rangle|/\tilde{q}_{tot}$ represents the fraction of the advective scalar flux in the total scalar flux. For comparison purpose, the curves of scalar transport with pure diffusion are also included. The results of pure diffusion were obtained by simulating the scalar transport in a quiescent fluid using the same numerical method. For both $Re_S$, the variations of $\langle\phi\rangle$, $\langle\partial\phi/\partial\tilde{z}\rangle$ and $|\langle\tilde{u}_z\phi\rangle|/\tilde{q}_{tot}$ with vertical coordinate $\tilde{z}$ demonstrate similar dependence on $Sc$. As $Sc$ increases, the advection effect gets stronger, and the curves deviate more from those of pure diffusion. For a given $Sc$, higher $Re_S$ leads to a stronger advection effect and more deviation amongst the curves. For pure diffusion, $\langle\phi\rangle$ is linearly dependent on $\tilde{z}$, that is, $\langle\partial\phi/\partial\tilde{z}\rangle$ is a constant above the herringbone structures to maintain a constant scalar flux in vertical direction. The advective scalar flux $\langle\tilde{u}_z\phi\rangle$ is zero everywhere. At nonzero Reynolds numbers, the disturbances produced by the herringbone structures cause a local mixing above the herringbone ridges, which not only decreases the vertical gradient of $\langle\phi\rangle$ in the local area, but also provides an advective scalar transfer mechanism. With the increase in $Sc$, the advection effect becomes stronger and the gradient of $\langle\phi\rangle$ becomes smaller.

For $Re_S = 20$, as shown in Fig. 7, with the increase in $Sc$ the curve of $\langle\phi\rangle$ becomes more vertical at about $\tilde{z} = 1.5$, and the curve of $\langle\partial\phi/\partial\tilde{z}\rangle$ curves leftward and forms a peak with the minimum $\langle\partial\phi/\partial\tilde{z}\rangle$. At the same time, the curve of $|\langle\tilde{u}_z\phi\rangle|/\tilde{q}_{tot}$ curves rightward and forms a peak with the maximum $|\langle\tilde{u}_z\phi\rangle|/\tilde{q}_{tot}$. This means that the fraction of advective scalar flux of the total scalar flux increases with $Sc$. At the top plane ($\tilde{z} = 3$), scalar transport occurs by pure diffusion because the vertical velocity is 0 there. When $Sc$ increases from 1 to 50, the curve slope of $\langle\phi\rangle$, that is, the value of $\langle\partial\phi/\partial\tilde{z}\rangle$ at the top plane, also increases. This means that the total scalar flux increases with $Sc$ as a result of enhanced advective scalar transfer. Since the passive scalar is absorbed at both the side walls of the herringbone ridges and the bottom surfaces, the curve of $\langle\phi\rangle$ is not a straight line in the grooves ($\tilde{z} < 1$) for pure diffusion. At $Re_S = 20$, the flow advection is weak in the grooves, so the curves of $\langle\phi\rangle$ and $\langle\partial\phi/\partial\tilde{z}\rangle$ do not change much with $Sc$ when $\tilde{z} < 1$, yet the curves of $|\langle\tilde{u}_z\phi\rangle|/\tilde{q}_{tot}$ still show the enhancement of advective scalar transport in the grooves.

For $Re_S = 200$, the curves shown in Fig. 8 demonstrate a strong advection effect due to the decrease in the kinematic viscosity of the fluid – except the curve for $Sc = 1$. As shown in the figure, the curves of $\langle\phi\rangle$, $\langle\partial\phi/\partial\tilde{z}\rangle$, and $|\langle\tilde{u}_z\phi\rangle|/\tilde{q}_{tot}$ for $Sc = 1$ almost coincide with those of pure diffusion. However, when $Sc$ increases to 10, the value of $\langle\phi\rangle$ decreases and the curve of $\langle\phi\rangle$ is almost vertical above the herringbone structures. Correspondingly, the curve of $\langle \partial \phi / \partial \tilde{z} \rangle$ has a minimum value close to 0 at about $\tilde{z} = 1.5$, and the curve of $|\langle \tilde{u}_z \phi \rangle| / \tilde{q}_{tot}$ shows a maximum value close to 1 at that point. This means that the advective scalar transfer has become the dominant means of scalar transport.

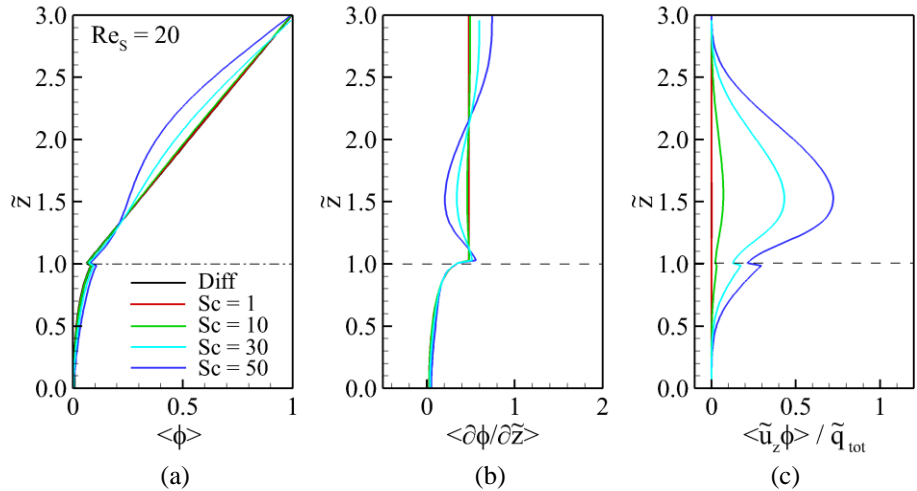

(a)          (b)          (c)

Fig. 7. Profiles of (a) horizontally averaged scalar concentration $\langle\phi\rangle$, (b) vertical gradient of scalar concentration $\langle\partial\phi/\partial\tilde{z}\rangle$, and (c) ratio of advective scalar flux to total scalar flux $|\langle\tilde{u}_z\phi\rangle|/\tilde{q}_{tot}$ at $Re_S = 20$.

As $Sc$ increases from 10 to 50, the value of $\langle \phi \rangle$ further decreases above the herringbone structures, and the area of small $\langle \partial \phi / \partial \tilde{z} \rangle$ (close to 0) expands both upward and downward. Correspondingly, the area of large $|\langle \tilde{u}_z \phi \rangle| / \tilde{q}_{tot}$ (close to 1) also expands. The advective scalar transfer dominates in the majority of the upper flow. At the top plane, scalar transport still relies on pure diffusion. As $Sc$ increases from 1 to 50, the curve slope of $\langle \phi \rangle$, that is, the value of $\langle \partial \phi / \partial \tilde{z} \rangle$ at the top plane, also increases indicating an increase in total scalar flux from the top plane to the flow. In the grooves, $\langle \phi \rangle$ smoothly decreases from a small value at ridge tips to 0 at the bottom surface for pure diffusion, and the curve slope of $\langle \phi \rangle$, that is, the value of $\langle \partial \phi / \partial \tilde{z} \rangle$ is close to 0 at the bottom surface. This suggests that only a small fraction of scalar is transported to the bottom surface through pure diffusion. At $Sc = 1$, the scalar transfer characteristics are almost the same as those of pure diffusion. When $Sc$ increases to 10, the curve of $\langle \phi \rangle$ becomes vertical below the ridge tips, and the value of $\langle \partial \phi / \partial \tilde{z} \rangle$ decreases toward 0. Correspondingly, the curve of $|\langle \tilde{u}_z \phi \rangle| / \tilde{q}_{tot}$ shows apparent increase in $|\langle \tilde{u}_z \phi \rangle| / \tilde{q}_{tot}$ in that area. As $Sc$ further increases, the flow recirculation in the grooves becomes stronger, further decreasing the value of $\langle \partial \phi / \partial \tilde{z} \rangle$ and increasing the value of $|\langle \tilde{u}_z \phi \rangle| / \tilde{q}_{tot}$ below the ridge tips. At the bottom surface ($\tilde{z} = 0$), the increase in $Sc$ leads to the increase in $\langle \partial \phi / \partial \tilde{z} \rangle$, i.e., the scalar flux transported to the bottom surface increases.

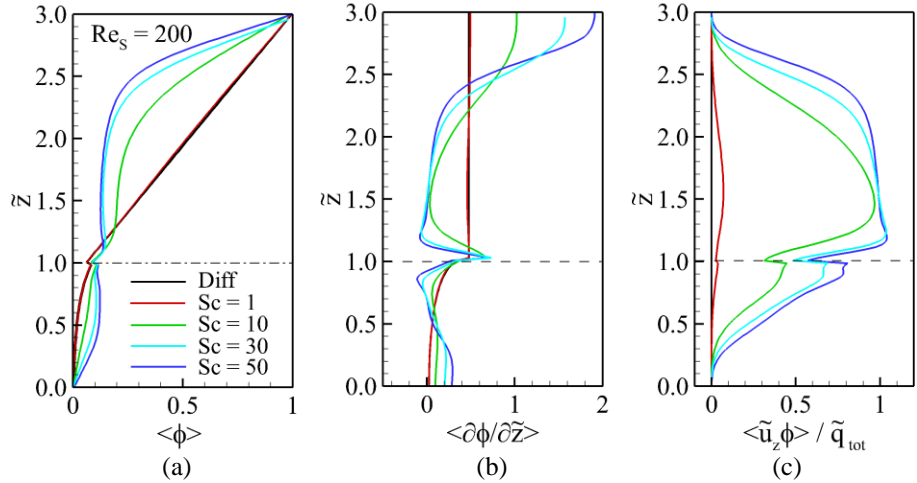

Fig. 8. Profiles of (a) horizontally averaged scalar concentration $\langle \phi \rangle$, (b) vertical gradient of scalar concentration $\langle \partial \phi / \partial \tilde{z} \rangle$, and (c) ratio of advective scalar flux to total scalar flux $|\langle \tilde{u}_z \phi \rangle| / \tilde{q}_{tot}$ at $Re_S = 200$.

To measure the enhancement of scalar transport from the top plane to the herringbone surfaces and bottom plane, we compare the total scalar flux in the presence of herringbone structures with that of a smooth bottom plane. The scalar transport between smooth planes is performed by pure diffusion for laminar flows. We select two distances $H_{smo}$ between the top and bottom smooth planes. For the first one, $\widetilde{H}_{smo} = 3\tilde{h}$, corresponding to the case in which the top and bottom planes are at the same positions as those of the cases with herringbone structures. For the second one, $\widetilde{H}_{smo} = 2\tilde{h}$, which corresponds to the case in which the top plane is at the same position as that considered in this study, yet the bottom plane is placed at the tips of the herringbone structures. The nondimensional scalar fluxes between the smooth planes are calculated as,

$$\tilde{q}_{smo} = (\phi_0 - 0)/\widetilde{H}_{smo} = 1/3 \quad \text{when} \quad \widetilde{H}_{smo} = 3\tilde{h} \tag{18}$$

$$\tilde{q}_{smo} = (\phi_0 - 0)/\widetilde{H}_{smo} = 1/2 \quad \text{when} \quad \widetilde{H}_{smo} = 2\tilde{h} \tag{19}$$

Figure 9 shows the ratio of total scalar flux with herringbone structures to those with the smooth bottom wall $\tilde{q}_{tot}/\tilde{q}_{smo}$ versus Schmidt number $Sc$ in logarithm scales. Compared with the smooth planes with $\widetilde{H}_{smo} = 3\tilde{h}$, the herringbone structures significantly enhance the scalar transport by reducing the transfer distance, increasing surface area, and inducing flow advection in vertical direction. As a result, $\tilde{q}_{tot}/\tilde{q}_{smo}$ is always greater than 1, as shown in Fig. 9(a).

For pure diffusion, which is equivalent to $Re_S = 0$, $\tilde{q}_{tot}/\tilde{q}_{smo}$ denoted as $(\tilde{q}_{tot}/\tilde{q}_{smo})_{dif}$, is independent of $Sc$ and remains at a value about 1.43. This value is determined by the distance between the top and bottom planes and the geometric parameters of the herringbone structures. For $Re_S > 0$, $\tilde{q}_{tot}/\tilde{q}_{smo}$ increases with both $Sc$ and $Re_S$ from the value of pure diffusion. At $Re_S = 20$, the scalar transport is dominated by diffusion from $Sc = 0$ to a critical Schmidt number $Sc_{crit}$ around 10, after which $\tilde{q}_{tot}/\tilde{q}_{smo}$ increases monotonically with $Sc$ due to the enhancement from flow advection. It is apparent that $Sc_{crit}$ decreases with the increase in $Re_S$. At $Re_S = 100$ and 200, $\tilde{q}_{tot}/\tilde{q}_{smo}$ is close to $(\tilde{q}_{tot}/\tilde{q}_{smo})_{dif}$ when $Sc = 1$. After that, $\tilde{q}_{tot}/\tilde{q}_{smo}$ increases remarkably with the increase in $Sc$. Note that in logarithm scale the curves appear as straight lines and the slopes are almost the same when $Sc \geq 10$ for $Re_S = 100$ and 200. This implies the variation of $\tilde{q}_{tot}/\tilde{q}_{smo}$ follows a power law on $Sc$ when advective scalar transport dominates, and the power of $Sc$ is independent of $Re_S$.

For the flows considered in this study, it is found that the power approaches eventually to a value close to 0.4 as $Sc$ increases. Figure 9(b) shows the enhancement of scalar transfer by herringbone structures compared with the flow over smooth planes with $\widetilde{H}_{smo} = 2\tilde{h}$. In this comparison, the herringbone structures increase the scalar transport distance from the top plane to the herringbone surfaces and bottom planes, so $\tilde{q}_{tot}/\tilde{q}_{smo}$ becomes less than 1 for pure diffusion. However, with the increase in $Sc$ and $Re_S$, the strengthened advection effect not only offsets the reduction of scalar transfer caused by increased transport distance, but also causes a significant increase in $\tilde{q}_{tot}/\tilde{q}_{smo}$. Since $\tilde{q}_{smo}$ is independent of $Sc$ and $Re_S$, the enhancement of scalar transport, $\tilde{q}_{tot}/\tilde{q}_{smo}$, can be written in a unified form,

$$\tilde{q}_{tot}/\tilde{q}_{smo} = f(Re_S, Sc, \boldsymbol{\gamma}) \tag{20}$$

where $\boldsymbol{\gamma}$ is a vector of the geometric parameters, , such as the distance between the two planes and the height of herringbone structures, and $f(Re_S, Sc, \boldsymbol{\gamma})$ is a function of $Re_S$, $Sc$ and $\boldsymbol{\gamma}$. The variation of $\tilde{q}_{tot}/\tilde{q}_{smo}$ with $Sc$ shown in Fig. 9 suggests that $\tilde{q}_{tot}/\tilde{q}_{smo}$ has a power law relationship with $Sc$ with a power of 0.4 when advective scalar transport takes effect. Therefore, Eqn. (20) can be further written as,

$$\tilde{q}_{tot}/\tilde{q}_{smo} = g(Re_S, \boldsymbol{\gamma})Sc^{0.4} \tag{21}$$

where $g(Re_S, \boldsymbol{\gamma})$ is a function of $Re_S$ and $\boldsymbol{\gamma}$. For turbulent heat transfer, it has been found that the Nusselt number $Nu$, a nondimensional parameter characterizing the heat transfer rate, can be expressed as the famous Dittus-Boelter equation [30],

$$Nu = 0.023Re^{0.8}Pr^{0.4} \tag{22}$$

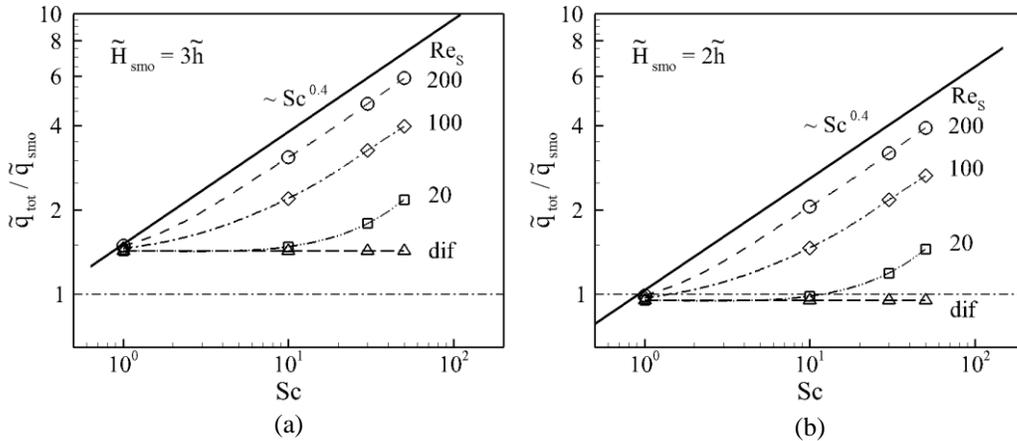

Fig. 9. Ratio of total scalar flux with herringbone structures to that with smooth bottom wall versus Schmidt number. (a) $\widetilde{H}_{smo} = 3\tilde{h}$, and (b) $\widetilde{H}_{smo} = 2\tilde{h}$. 'dif' indicates the cases of pure diffusion.

In this expression, the power of $Pr$ is 0.4, which is the same value as that disclosed in Eqn. (21).

To further explore the dependence of scalar transfer rate on $Re_S$ and $Sc$, we selected the data points at $Sc = 10, 30$ and $50$ in Fig. 9(a) and plot $\tilde{q}_{tot}/\tilde{q}_{smo}$ versus $Re_S$ in logarithm scale in Fig. 10. To compare with the Dittus-Boelter equation, we also plot the line of $\tilde{q}_{tot}/\tilde{q}_{smo} \propto Re^{0.8}$. As shown in the figure, the curves of three Schmidt numbers exhibit almost the same slopes at every $Re_S$ except in the lower left region around the point at $Re_S = 20$ and $Sc = 10$, where the scalar transport is dominated by diffusion. The curve slopes asymptotically approach the slope of $\tilde{q}_{tot}/\tilde{q}_{smo} \propto Re^{0.8}$ as $Re_S$ increases. However, the finite range of $Re_S$ considered in this study cannot tell if the power of $Re_S$ will finally become 0.8. It is almost certain that the dependence of scalar transport on $Sc$ will become a power law as $Re_S$ further increases, and the power is close to 0.8. Once the power law relationship is reached, the expression of $\tilde{q}_{tot}/\tilde{q}_{smo}$ given in Eqn. (21) can be further decomposed as,

$$\tilde{q}_{tot}/\tilde{q}_{smo} = h(\boldsymbol{\gamma})Re_S^m Sc^{0.4} \tag{23}$$

where $m$ is a value close to 0.8. This expression suggests that when $Re_S$ is large enough, the heat and mass transfer rates also have a power law relationship with $Re_S$, and the effects of geometric parameters $\boldsymbol{\gamma}$, Reynolds number $Re_S$, and Schmidt number $Sc$ are decoupled. The scalar transfer rate can be expressed in the general form as,

$$\tilde{q}_{tot} = f(Re_S, Sc, \boldsymbol{\gamma}) \xrightarrow{adv.\ dom} g(Re_S, \boldsymbol{\gamma})Sc^{0.4} \xrightarrow{large\ Re_S} c(\boldsymbol{\gamma})Re_S^m Sc^{0.4} \tag{24}$$

where $m \approx 0.8$, and 'adv. dom.' represents advection dominant.

If the scalar transfer rates from the numerical simulations are fitted within the finite range of $Re_S$ to the power law relationship with the least square method, there is obtained,

$$\tilde{q}_{tot} = c(\boldsymbol{\gamma})Re_S^{0.42} Sc^{0.4} \tag{25}$$

The power of $Re_S$ is 0.42, smaller than that in the Dittus-Boelter equation.

The Peclet number, which is defined as the product of Reynolds and Schmidt numbers, is also an important nondimensional parameter characterizing scalar transport. Figure 11 shows the enhancement of scalar transfer $\tilde{q}_{tot}/\tilde{q}_{smo}$ versus Peclet number $Pe$ for two $\widetilde{H}_{smo}$. As shown in the figure, all the curves demonstrate a power law relationship with $Pe$ in the logarithm scale when $Pe$ increases above a critical value around 200. The curves at different values of $Re_S$ group together and form a dense cluster of curves. The curve slopes are roughly the same as that of $Pe^{0.4}$. If we use Eqn. (24) to approximate the scalar transport rate in the range considered in this study, the expression can be rearranged as

$$\tilde{q}_{tot} = c(\boldsymbol{\gamma})Re_S^{0.02}Pe^{0.4} \tag{26}$$

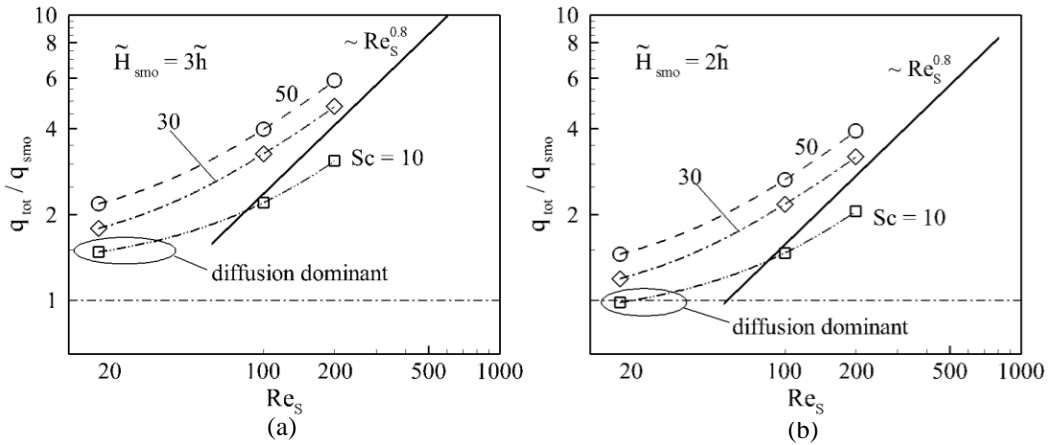

Fig. 10. Ratio of total scalar flux with herringbone structures to that with smooth bottom wall versus Reynolds number. (a) $\widetilde{H}_{smo} = 3\tilde{h}$, and (b) $\widetilde{H}_{smo} = 2\tilde{h}$.

The term $Re_S^{0.02}$, that is, the Reynolds number effect causes the deviation among the curves at different $Re_S$ values. However, for the $Re_S$ in this study, $Re_S^{0.02} \approx 1$, which makes the curves close to each other. Therefore, the scalar transport rate can be roughly estimated by

$$\tilde{q}_{tot} = c(\boldsymbol{\gamma})Pe^{0.4} \tag{27}$$

when $Re_S$ is not very large. This simplified expression significantly facilitates the estimation of heat and mass transfer rate in industrial applications.

It needs to be pointed out that the mechanisms and conclusions made in this paper can also be applied to the cases with heat and mass transferred from the bottom surfaces to the top plane after a simple conversion.

## 5. Conclusion

Focusing on the characterization of a simple shear flow over staggered herringbone structures, we have identified the mechanisms of heat and mass transfer from the top plane to the herringbone surfaces and bottom plane, induced by the transverse flow caused by the repetitive herringbone structures, and have found out the dependence of heat and mass transfer characteristics on Reynolds and Schmidt numbers.

Through high-fidelity numerical simulation, two basic flow motions were identified, which provide the fundamental mechanisms for heat and mass transfer through flow advection. The first is the spiral flow oscillation between the top plane and the herringbone structures, which transports heat and mass from the top plane (or bulk flow) to the regions around herringbone tips. The second is the flow recirculation in the grooves between herringbone ridges, which transports heat and mass from the areas around herringbone tips to the side walls of herringbone ridges and the bottom surfaces. These two basic flow motions couple together to form the complex transport mechanisms.

When advective heat and mass transfer takes effect at relatively larger Reynolds and Schmidt numbers, the dependence of total transfer rate on the Schmidt number follows a power law and the power is the same as that in the well-known Dittus-Boelter equation for turbulent heat transfer. As the Reynolds number increases, the dependence of total transfer rate on the Reynolds number also approaches a power law, and the power is close to that in the Dittus-Boelter equation. When the power law on Reynolds number is reached, the dependence on geometric parameters, Reynolds numbers and Schmidt numbers can be decoupled. This discovery helps the development of physics-based models for a wide range of fundamental research and industrial applications.

## References

[1]. C.-M. Ho, and Y.-C. Tai, Micro-electro-mechanical-systems (MEMS) and fluid flows," Annu. Rev. Fluid Mech. 30 (1998) 579-612.

[2]. H. A. Stone, A. D. Stroock, and A. Ajdari, Engineering Flows in Small Devices: Microfluidics Toward a Lab-on-a-Chip, Annu. Rev. Fluid Mech. 36 (2004) 381-411.

[3]. D. T. Chiu, A. J. deMello, D. Di Carlo, P. S. Doyle, C. Hansen, R. M. Maceiczyk, and R. C. R. Wootton, Small but Perfectly Formed? Successes, Challenges, and Opportunities for Microfluidics in the Chemical and Biological Sciences, Chem. 2 (2017) 201-223.

[4]. A. D. Stroock, S. K. W. Dertinger, A. Ajdari, I. Mezić, H. A. Stone, and G. M. Whitesides, Chaotic Mixer for Microchannels, Sci. 295 (2002) 647-651.

[5]. J. M. Ottino, and S. Wiggins, Designing Optimal Micromixers, Sci. 305 (2004) 485-486.

[6]. J. T. Yang, K. J. Huang, and Y. C. Lin, Geometric effects on fluid mixing in passive grooved micromixers, Lab Chip. 5 (2005) 1140-1147.

[7]. J. Aubin, D. F. Fletcher, and C. Xuereb, Design of micromixers using CFD modelling, Chem. Eng. Sci. 60 (2005) 2503-2516.

[8]. D. G. Hassell, and W. B. Zimmerman, Investigation of the convective motion through a staggered herringbone micromixer at low Reynolds number flow, Chem. Eng. Sci. 61 (2006) 2977-2985.

[9]. S. Hardt, K. S. Drese, V. Hessel, and F. Schönfeld, Passive micromixers for applications in the microreactor and µTAS fields, Microfluidics and Nanofluidics 1 (2005) 108-118.

[10]. S. L. Stott, C.-H. Hsu, D. I. Tsukrov, M. Yu, D. T. Miyamoto, B. A. Waltman, S. M. Rothenberg, A. M. Shah, M. E. Smas, G. K. Korir, F. P. Floyd, A. J. Gilman, J. B. Lord, D. Winokur, S. Springer, D. Irimia, S. Nagrath, L. V. Sequist, R. J. Lee, K. J. Isselbacher, S. Maheswaran, D. A. Haber, and M. Toner, Isolation of circulating tumor cells using a microvortex-generating herringbone-chip, PNAS 107 (2010) 18392-18397.

[11]. M. A. Ansari, and K.-Y. Kim, Shape optimization of a micromixer with staggered herringbone groove, Chem. Eng. Sci. 62 (2007) 6687-6695.

[12]. C. A. Cortes-Quiroz, M. Zangeneh, and A. Goto, On multi-objective optimization of geometry of staggered herringbone micromixer, Microfluidics and Nanofluidics **7** (2009) 29-43.

[13]. S. Hossain, A. Husain, and K.-Y. Kim, Optimization of Micromixer with Staggered Herringbone Grooves on Top and Bottom Walls, Eng. Appl. Comp. Fluid Mech.5 (2011) 506-516.

[14]. P. B. Howell, Jr., D. R. Mott, S. Fertig, C. R. Kaplan, J. P. Golden, E. S. Oran, and F. S. Ligler, A microfluidic mixer with grooves placed on the top and bottom of the channel, Lab Chip. **5** (2005) 524-530.

[15]. D. Lin, F. He, Y. Liao, J. Lin, C. Liu, J. Song, and Y. Cheng, Three-dimensional staggered herringbone mixer fabricated by femtosecond laser direct writing, J. Opt. 15 (2013) 025601.

[16]. D. Gobby, P. Angeli, and A. Gavriilidis, Mixing characteristics of T-type microfluidic mixers, Journal of Micromech. Microeng. 11 (2001) 126-132.

[17]. F. Yang, M. Alwazzan, W. Li, and C. Li, Single- and Two-Phase Thermal Transport in Microchannels With Embedded Staggered Herringbone Mixers," J. Microelectro. Sys. 23 (2014) 1346-1358.

[18]. J. Marschewski, R. Brechbühler, S. Jung, P. Ruch, B. Michel, and D. Poulikakos, Significant heat transfer enhancement in microchannels with herringbone-inspired microstructures, Int. J. Heat Mass Trans. 95 (2016) 755-764.

[19]. Y. H. Qian, D. D'Humières, and P. Lallemand, Lattice BGK Models for Navier-Stokes Equation, Europhys. Lett. 17 (1992) 479-484.

[20]. S. Chen, and G. D. Doolen, Lattice Boltzmann method for fluid flows, Annu. Rev. Fluid Mech. 30 (1998) 329-364.

[21]. Y. Wang, J. Brasseur, G. Banco, A. Webb, A. Ailiani, and T. Neuberger, Development of a lattice-Boltzmann method for multiscale transport and absorption with application to intestinal function, Computational Modeling in Biomechanics, 69-96, Springer, 2010.

[22]. P. Lallemand, and L.-S. Luo, Lattice Boltzmann method for moving boundaries, J. Comp. Phys. 184 (2003) 406-421.

[23]. A. J. C. Ladd, Numerical simulations of particulate suspensions via a discretized Boltzmann equation. Part 1. Theoretical foundation, Journal of Fluid Mechanics 271 (1994) 285-309.

[24]. D. Frenkel, and M. H. Ernst, Simulation of diffusion in a two-dimensional lattice-gas cellular automaton: A test of mode-coupling theory, Phys. Rev. Lett. 63 (1989) 2165-2168.

[25]. C. P. Lowe, and D. Frenkel, The super long-time decay of velocity fluctuations in a two-dimensional fluid, Physica A. 220 (1995) 251-260.

[26]. R. M. H. Merks, A. G. Hoekstra, and P. M. A. Sloot, The Moment Propagation Method for Advection–Diffusion in the Lattice Boltzmann Method: Validation and Péclet Number Limits, J. Comp. Phys. 183 (2002) 563-576.

[27]. Y. Wang, J. G. Brasseur, G. G. Banco, A. G. Webb, A. C. Ailiani, and T. Neuberger, A multiscale lattice Boltzmann model of macro- to micro-scale transport, with applications to gut function, Phil. Trans. R. Soc. A 368 (2010) 2863-2880.

[28]. Y. Wang, and J. G. Brasseur, Three-dimensional mechanisms of macro-to-micro-scale transport and absorption enhancement by gut villi motions, Phys. Rev. E 95 (2017) 062412.

[29]. Y. Wang, and J. G. Brasseur, Enhancement of mass transfer from particles by local shear-rate and correlations with application to drug dissolution, AIChE J. 65 (2019) e16617.

[30]. H. W. McAdams, Heat Transmission (McGraw-Hill, 1954).